\documentclass{jkas}

\def\year{2022} % publication year
\def\month{December} % publication month
\def\volume{55} % journal volume
\def\issue{6} % journal issue
\def\beginpage{1} % first page of article
\def\endpage{} % last page of article
\def\received{September 16, 2022} % date paper was received by JKAS
\def\accepted{November 22, 2022} % date of acceptance
\date{Received \received ; accepted \accepted}

\usepackage{flushend} %% balance columns on last page
\usepackage[colorlinks,
            breaklinks,
            linkcolor=blue,
            urlcolor=blue,
            anchorcolor=blue,
            menucolor=blue,
            citecolor=blue]{hyperref}
\def\arcsec{\hbox{$^{\prime\prime}$}}

\def\farcm{\hbox{$.\mkern-4mu^\prime$}}
\def\farcs{\hbox{$.\!\!^{\prime\prime}$}}
\title{
Infrared Properties of OGLE4 Mira variables in our Galaxy
}

\author{\href{https://orcid.org/0000-0001-9104-9763}{Kyung-Won Suh}}
\affil[]
{Department of Astronomy and Space Science, Chungbuk National University,
Chungbuk, Cheongju-City 28644, Korea; \email{kwsuh@chungbuk.ac.kr}}
\def\corrauthor{
K.-W. Suh
}

\def\runningauthor{
Suh
}

\def\runningtitle{
Infrared Properties of OGLE4 Mira variables in our Galaxy
}

\def\keywords{
{Long Period Variables --- stars: AGB and post-AGB --- circumstellar matter
--- infrared: stars --- radiative transfer}
}

\def\abstracttext{
We investigate infrared properties of OGLE4 Mira variables in our Galaxy. For
each object, we cross-identify the AllWISE, 2MASS, Gaia, and IRAS counterparts.
We present various IR two-color diagrams (2CDs) and period-magnitude and
period-color relations for the Mira variables. Generally, the Mira variables with
longer periods are brighter in the IR fluxes and redder in the IR colors. In this
work, we also revise and update the previous catalog of AGB stars in our Galaxy
using the new sample of OGLE4 Mira variables. Now, we present a new catalog of
74,093 (64,609 O-rich and 9484 C-rich) AGB stars in our Galaxy. A group of 23,314
(19,196 O-rich and 4118 C-rich) AGB stars are identified based on the IRAS PSC
and another group of 50,779 (45,413 O-rich and 5366 C-rich) AGB stars are
identified based on the AllWISE source catalog. For all of the AGB stars, we
cross-identify the IRAS, AKARI, MSX, AllWISE, 2MASS, OGLE4, Gaia, and AAVSO
counterparts and present various infrared 2CDs. Comparing the observations with
the theory, we find that basic theoretical dust shell models can account for the
IR observations fairly well for most of the AGB stars.
}
\begin{document}
\jkashead %% set title, authors, abstract, etc.

%%%%%%%%%%%%%%%%%%%%%%%%%%%%%%%%%%%%%%%%%%%%%%%%%%%%%%%%%%%%%%%%%%%%%
%%% BEGIN MAIN TEXT HERE %%%%%%%%%%%%%%%%%%%%%%%%%%%%%%%%%%%%%%%%%%%%
%%%%%%%%%%%%%%%%%%%%%%%%%%%%%%%%%%%%%%%%%%%%%%%%%%%%%%%%%%%%%%%%%%%%%

\section{Introduction\label{sec:intro}}

Most asymptotic giant branch (AGB) stars are believed to be long-period variables
(LPVs) with outer dust envelopes (e.g., \citealt{suh2021}). LPVs are classified
into small-amplitude red giants (SARGs), semiregular variables (SRVs), and Mira
variables (e.g., \citealt{swu13}). Though most of SRVs and a majority of SARGs
could also be in the AGB phase, all Mira variables are known to be in the AGB
phase (e.g., \citealt{hofner2018}).

Depending on the chemical composition of the photosphere or the outer dust
envelope, AGB stars are classified into O-rich or C-rich stars. Amorphous
silicates in the envelopes around O-rich AGB (OAGB) stars produce 10 $\mu$m and
18 $\mu$m features in emission or absorption (e.g., \citealt{suh1999}). Dust
envelopes around C-rich AGB (CAGB) stars consist of amorphous carbon (AMC) and
SiC dust (e.g., \citealt{suh2000}). SiC dust produces the 11.3 $\mu$m emission
feature.

\citet{suh2021} presented a catalog of 11,209 OAGB stars and 7172 CAGB stars in
our Galaxy and presented various IR two-color diagrams (2CDs) for the AGB stars.
The IR 2CDs have been useful in studying the chemical and physical properties of
various celestial objects (e.g., \citealt{vanderveen1988}; \citealt{suh2015}).

Thanks to the optical gravitational lensing experiment IV (OGLE4) project, the
number of known Mira variable in our Galaxy has increased significantly.
\citet{iwanek2022} presented a sample of 65,981 OGLE4 Mira variables in our
Galaxy, from which the majority (47,532 objects) comprises new discoveries.

In this work, we investigate IR properties of OGLE4 Mira variables in our Galaxy.
For each object, we cross-identify the Wide-field Infrared Survey Explorer
(WISE), Two-Micron All-Sky Survey (2MASS), and Infrared Astronomical Satellite
(IRAS) counterparts. We present period-magnitude and period-color relations for
the Mira variables. We also revise and update the previous catalog of AGB stars
in our Galaxy using the new sample of OGLE4 Mira variables. For the new list of
AGB stars, we cross-identify the IRAS, AKARI, Midcourse Space Experiment (MSX),
AllWISE, 2MASS, OGLE4, Gaia Data Release 3 (DR3), and American Association of
Variable Star Observers (AAVSO; international variable star index; version 2022
September 23; \citealt{watson2022}) counterparts and present various IR 2CDs. We
compare the observations with theoretical models and discuss general properties
of the AGB stars.

\begin{table*}
\centering
\caption{OGLE4 Mira variables in our Galaxy\label{tab:tab1}}
\begin{tabular}{lllllllll}
\toprule
OGLE4M$^1$ & AllWISE$^2$ & 2MASS$^2$ & Gaia$^{2,3}$ & IRAS$^4$\\
\midrule
65,981  & 64,962 & 65,931 & 57,482 & 16,379 &  &    \\
\bottomrule
\end{tabular}
\tabnote{
$^1$OGLE4 Mira variables from \citet{iwanek2022}.
$^2$The number of counterparts within 5$\arcsec$ (2$\arcsec$ for Gaia; duplicate sources are excluded).
$^3$ Gaia DR3 LPV sources (\citealt{lebzelter2022}).
$^4$The number of positive counterparts of IRAS PSC sources within 60$\arcsec$ (see Section~\ref{sec:iras}).}
\end{table*}

\begin{figure*}[!t]
\centering
\includegraphics[width=122mm]{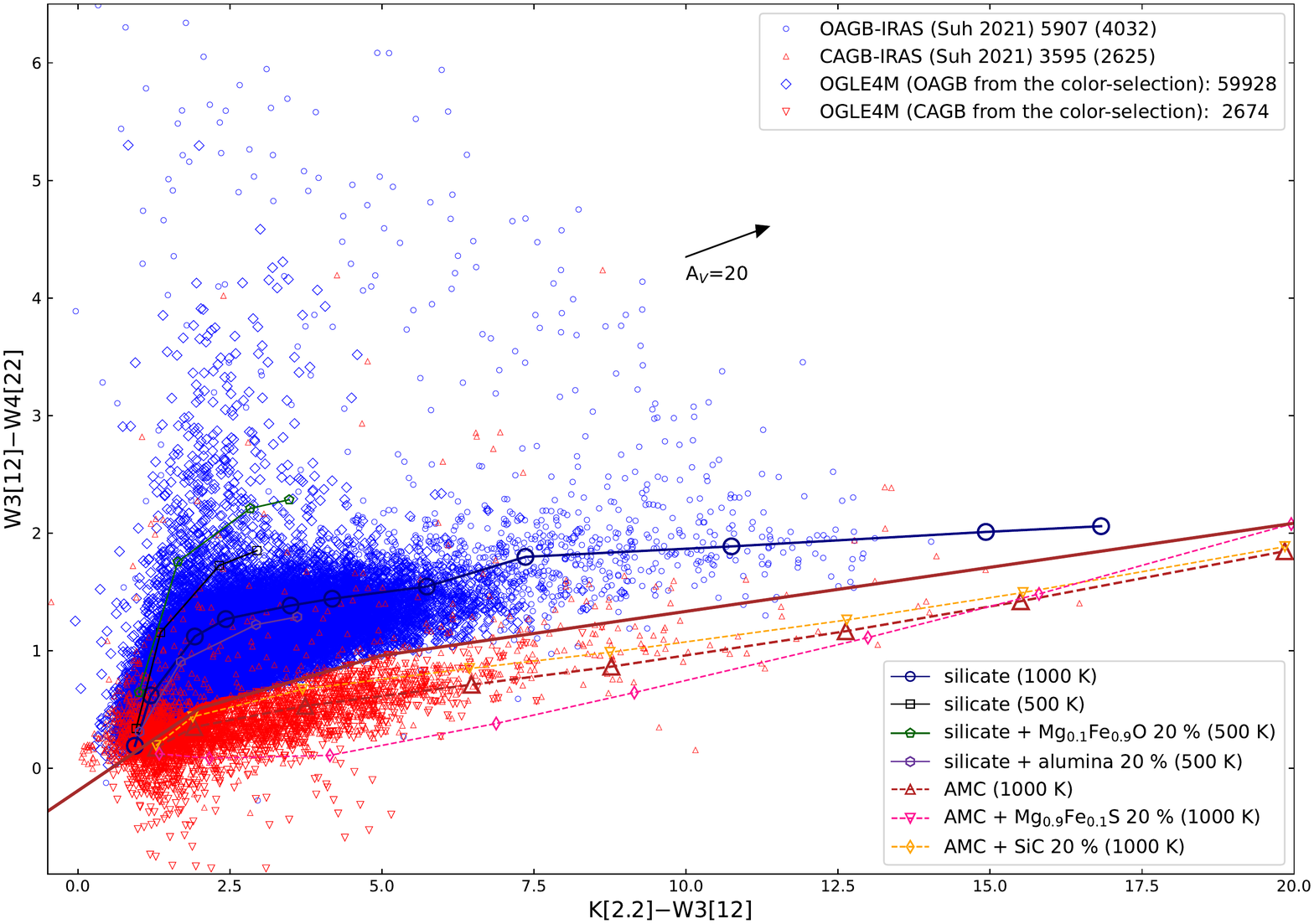} \\
\includegraphics[width=122mm]{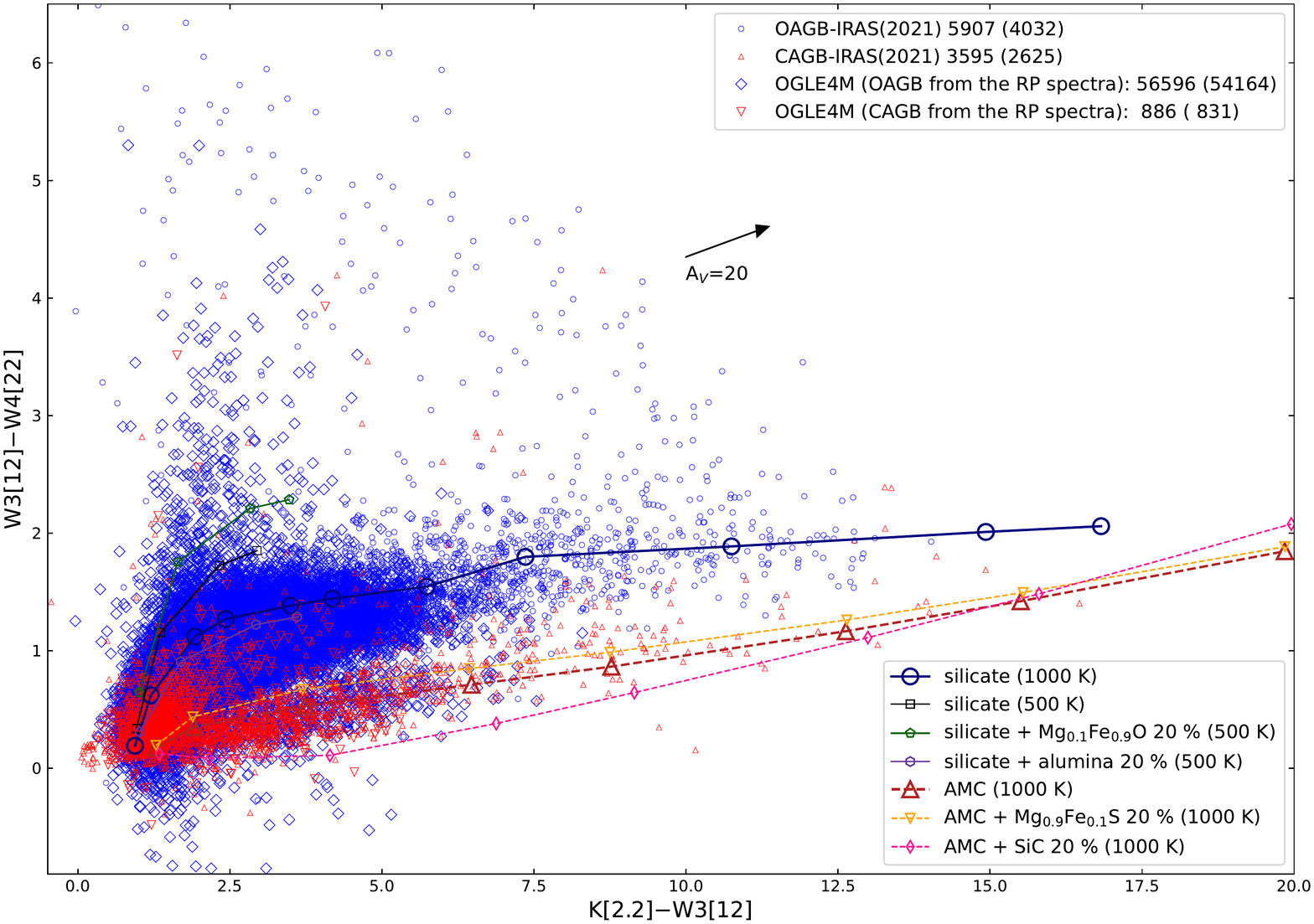}
\caption{WISE-2MASS 2CDs for color-selected OAGB and CAGB stars (the upper panel; see Section \ref{sec:w2cd}) and
for OAGB and CAGB stars classified from the Gaia RP spectra (the lowe panel; see Section \ref{sec:rpspectra}) for the OGLE4 Mira variables.
The 2CDs also show known AGB stars from \citet{suh2021} (see Table~\ref{tab:tab3}).
The observations are compared with theoretical models (see Section~\ref{sec:models}).
The brown line in the upper-panel shows the boundary line between OAGB and CAGB stars for the color-selection method.
The number of objects and the number of plotted objects (in parenthesis) with good-quality observed data are shown for each class.
\label{f1}}
\end{figure*}

\begin{figure}[!t]
\centering
\includegraphics[width=86mm]{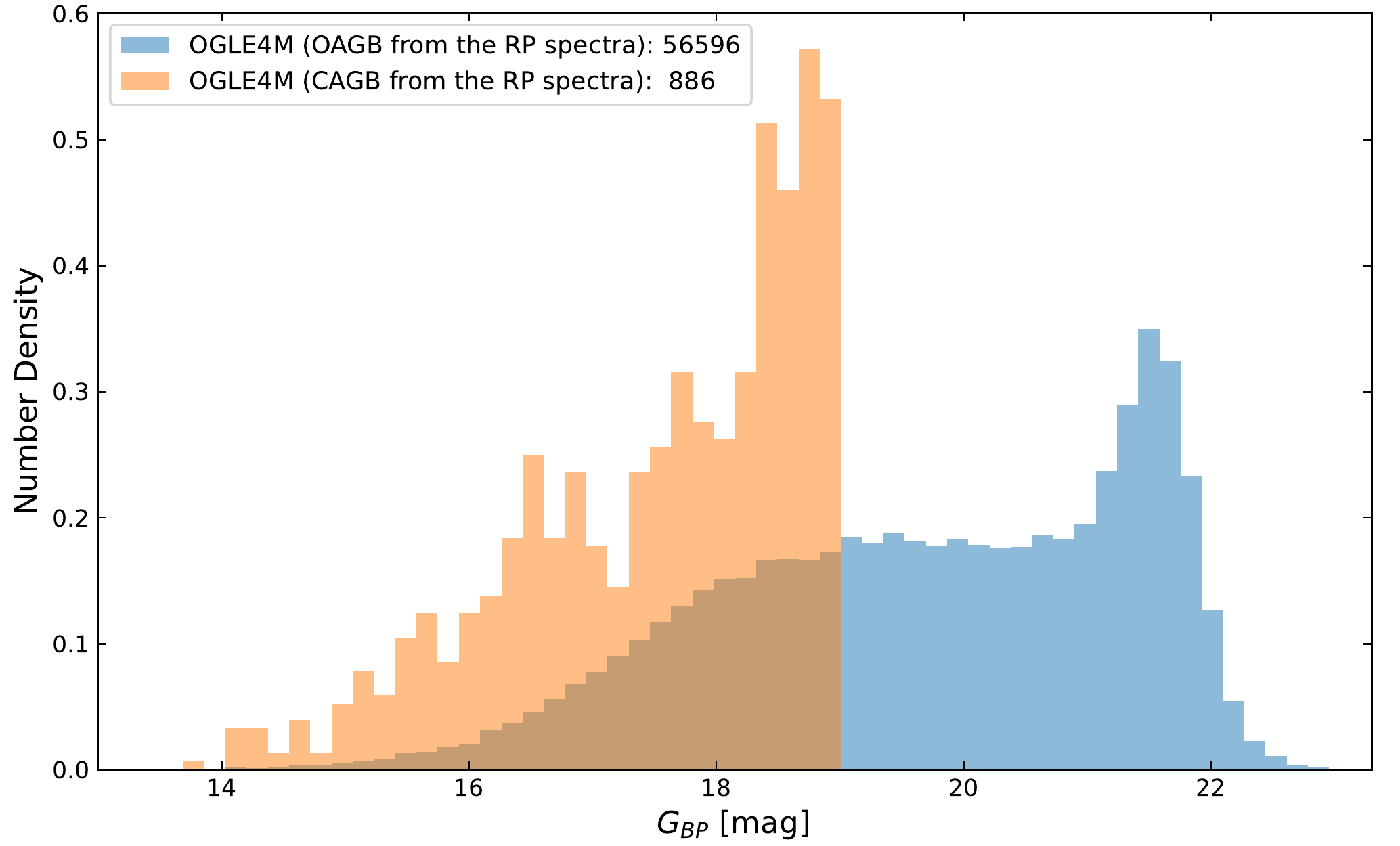}
\caption{Number density distributions of the Gaia magnitudes at the
BP band for OAGB and CAGB stars classified from the Gaia RP spectra for the
OGLE4 Mira variables (see Section~\ref{sec:rpspectra}).\label{f2}}
\end{figure}

\begin{figure}[!t]
\centering
\includegraphics[width=86mm]{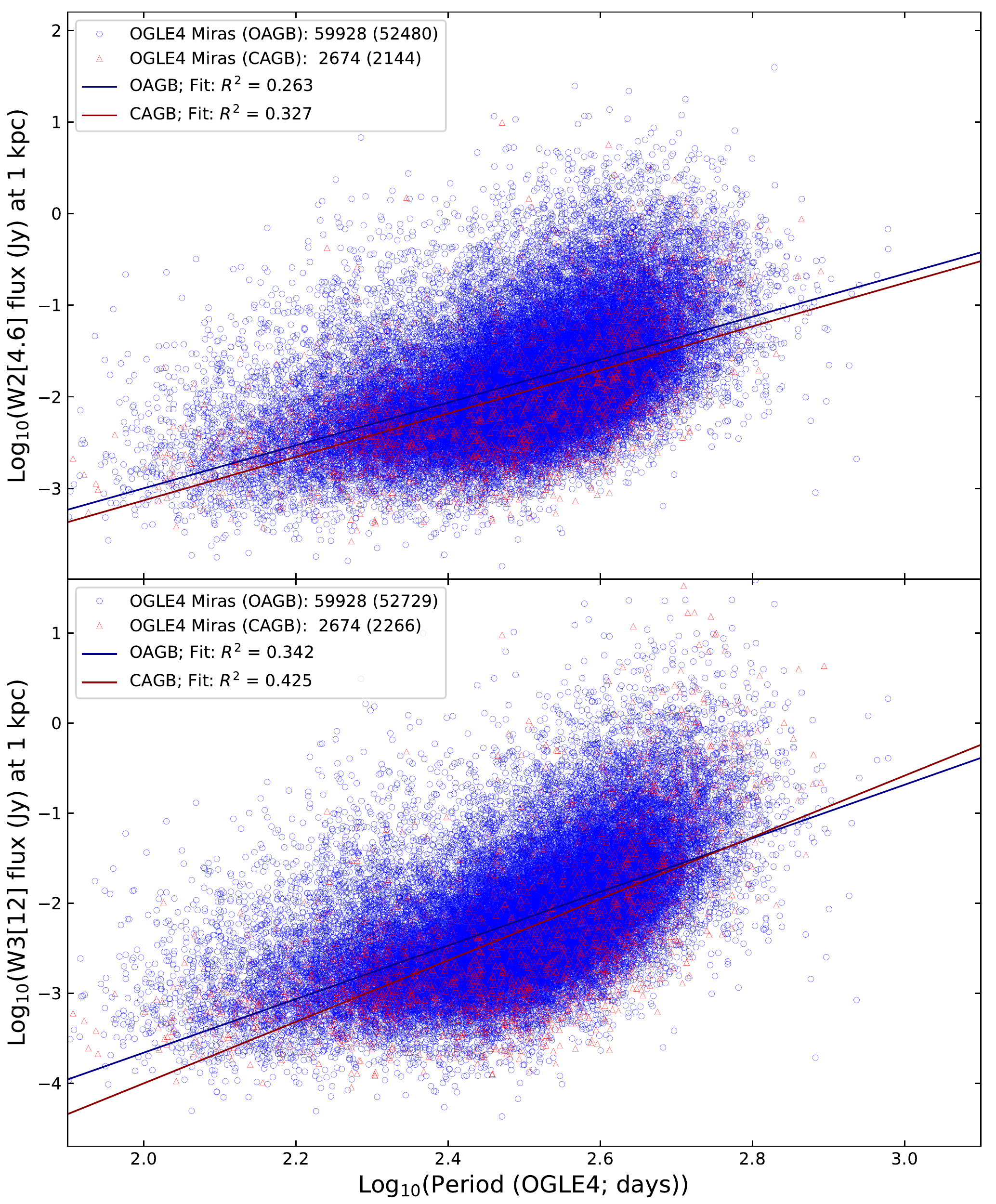}
\caption{Period-Magnitude relations for the OGLE4 Mira variables from
\citet{iwanek2022} with known distances from Gaia DR3 (see Section~\ref{sec:pmr}).\label{f3}}
\end{figure}

\begin{figure}[!t]
\centering
\includegraphics[width=86mm]{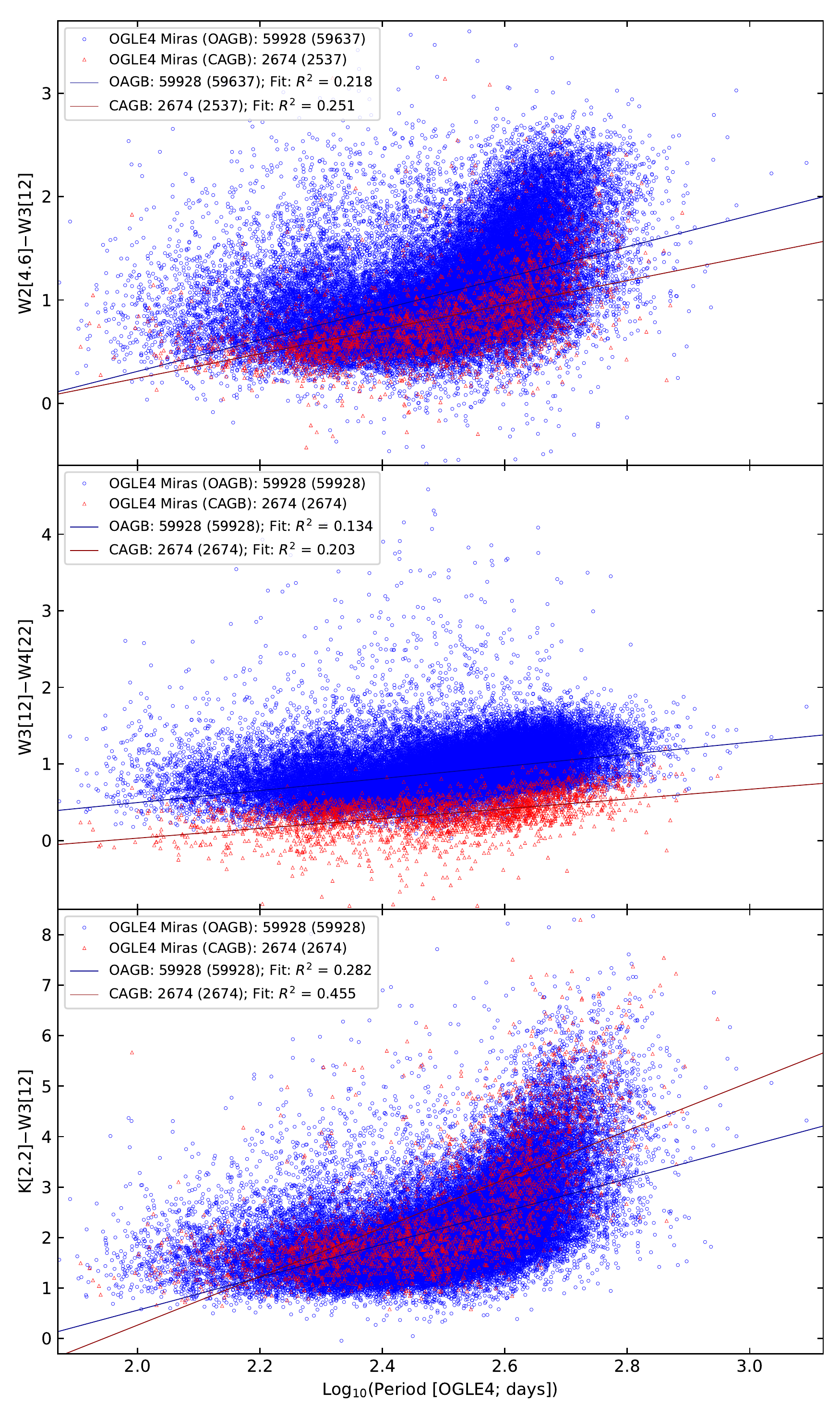}
\caption{Period-color relations for the OGLE4 Mira variables from
\citet{iwanek2022}.\label{f4}}
\end{figure}

\section{OGLE4 Mira variables in our Galaxy\label{sec:sample}}

\citet{iwanek2022} presented a sample of 65,981 OGLE4 Mira variables in our
Galaxy (40,356 objects in the Galactic bulge fields and 25,625 objects in the
Galactic disk).

\subsection{Infrared Photometric Data\label{sec:photdata}}

The IRAS point source catalog (PSC) \citep{beichman1988} provided photometric
data at four bands (12, 25, 60, and 100 $\mu$m) for 245,889 objects. The IRAS's
photometric sensitivity (averaged 10 sigma value at the 12 $\mu$m band) is 0.7
Jy. MSX (\citealt{egan2003}) conducted the Galactic plane survey with a higher
sensitivity (about 0.17 Jy at the 8.28 $\mu$m band) at four MIR bands (8.28,
12.13, 14.65, and 21.34 $\mu$m) for 454,091 objects (MSXC6). AKARI
\citep{murakami2007} provided the PSC data at two bands (9 and 18 $\mu$m;
\citealt{ishihara2010}) for 870,973 objects and bright-source catalogue (BSC)
data at four bands (65, 90, 140, and 160 $\mu$m; \citealt{yamamura2010}) for
427,071 objects. The field of view (FOV) pixel sizes of the IRAS, MSX, AKARI PSC,
and AKARI BSC images are 0$\farcm$75x(4$\farcm$5-4$\farcm$6), 18$\farcs$3,
10$\arcsec$, and 30$\arcsec$, respectively.

Two-Micron All-Sky Survey (2MASS) \citep{cutri2003} provided PSC data at J (1.25
$\mu$m), H (1.65 $\mu$m), and K (2.16 $\mu$m) bands for more than 470 million
objects. The FOV pixel size of the 2MASS image is 2$\arcsec$. WISE
\citep{wright2010} conducted the entire sky survey at W1 (3.4 $\mu$m), W2 (4.6
$\mu$m), W3 (12 $\mu$m), and W4 (22 $\mu$m) bands. For the four WISE bands, the
FOV pixel sizes are 2$\farcs$75, 2$\farcs$75, 2$\farcs$75, and 5$\farcs$5, and
the 5$\sigma$ photometric sensitivities are 0.068, 0.098, 0.86, and 5.4 mJy. In
2013, the AllWISE source catalog released to present more accurate positions and
four-band fluxes for over 747 million objects.

The FOV pixel size of the CCD camera system for OGLE4 is 0$\farcs$26 (see
\url{https://ogle.astrouw.edu.pl/}).

\subsection{Cross-matches\label{sec:crossmatch}}

Table~\ref{tab:tab1} lists the numbers of the cross-identified AllWISE, 2MASS,
Gaia, and IRAS PSC counterparts for the 65,981 OGLE4 Mira variables in our Galaxy
(\citealt{iwanek2022}). For each object, we find the nearest cross-matched source
within 5$\arcsec$ (2$\arcsec$ for Gaia). See Section~\ref{sec:iras} for the IRAS
counterparts.

In some cases, multiple OGLE4 objects may have the same cross-matched source
because the telescope for OGLE4 has a smaller FOV pixel size. So we have checked
all of the duplicated cross-matches and selected only the nearest object for the
one source in the AllWISE, 2MASS, Gaia, and IRAS PSC. We find that there are
64,962 OGLE4 Mira variables that have AllWISE counterparts.

We need to confirm that the cross-matched sources show the expected properties
(brightness, colors, etc) of the original object. A diagram comparing the fluxes
of the identified source and original source is useful. In most cases, they show
a good correlation (e.g., see Figure 1 in \citealt{suh2021}), but some objects
show large deviations. A reason for the deviations could be because they are
different objects. In this work, we have reduced the number of the deviated
objects by using the multiple cross-match and comparison processes (See
Section~\ref{sec:iras}). Because another reason for the deviations could be the
large scale variations of AGB stars or different beam sizes of the two telescope
systems, we do not exclude the remaining objects that show the large deviations.

IR 2CDs that show a general character of the object is also useful. For OGLE4
Mira variables, the IR 2CDs using the cross-matched sources need to show that the
new objects produce expected colors similar to previously known Mira variables or
AGB stars (Section~\ref{sec:w2cd}).

In this paper, we use only good-quality observational data for all wavelength
bands for the IRAS, 2MASS, WISE, AKARI, and MSX photometric data (q = 3 for IRAS
and AKARI; q = A for 2MASS; q = A or B for WISE; q = 3 or 4 for MSX).

\subsection{Color indices\label{sec:ircolor}}

The color index is defined by
\begin{equation}
M_{\lambda 1} - M_{\lambda 2} = - 2.5 \log_{10} {{F_{\lambda 1} / ZMF_{\lambda 1}} \over {F_{\lambda 2} / ZMF_{\lambda 2}}}
\end{equation}
where $ZMF_{\lambda i}$ is the zero magnitude flux (ZMF) at given wavelength
($\lambda i$). $ZMF_{\lambda i}$ is specified in the manual for the corresponding
telescope system (see Table 4 in \citealt{suh2021}).

Figures~\ref{f1}, \ref{f4}, and~\ref{f5} show IR 2CDs using various IR color
indices. Note that we use only good quality observational data for all wavelength
bands in plotting the IR 2CDs (see Section~\ref{sec:crossmatch}).

To consider the Galactic extinction processes, we plot reddening vectors for IR
2CD using NIR data (see \citealt{suh2021} for details of the reddening vectors).

\subsection{The 2MASS-WISE 2CD\label{sec:w2cd}}

IR 2CDs have been useful to distinguish different characteristics of various
heavenly bodies (e.g., \citealt{vanderveen1988}; \citealt{suh2018}). Though the
classification using the IR spectra would be much more reliable, the IR
color-selection method is useful because the method can make right classification
for a major portion of known OAGB or CAGB stars with IR spectroscopic data
(\citealt{sh2017}; \citealt{suh2021}). Therefore, the IR color-selection method
using an IR 2CD can be useful to distinguish between OAGB and CAGB for AGB stars
without reliable spectroscopic data.

The WISE-2MASS 2CD using W3[12]$-$W4[22] versus K[2.2]$-$W3[12] looks to be more
useful than any other 2CDs because the color-combination can distinguish
different classes of AGB stars more correctly (\citealt{suh2018};
\citealt{suh2020}; \citealt{suh2021}; see Section \ref{sec:agb-iras}) and the
good quality observed data (for the two IR colors) are available for a larger
number of sample objects. The brown line on the upper panel of Figure~\ref{f1}
shows the boundary line between OAGB and CAGB stars obtained for known AGB stars
(\citealt{suh2021}), which is used for the color-selection method in this paper.

The upper panel of Figure~\ref{f1} shows the WISE-2MASS 2CD for the OGLE4 Mira
variables (see Table~\ref{tab:tab1}) as well as the AGB stars in the catalog of
\citet{suh2021} (see Table~\ref{tab:tab3}). On this 2CD, we can find new
color-selected AGB stars from the 65,981 OGLE4 Mira variables: 59,928 OAGB and
2674 CAGB stars. Because another 3379 objects do not have the good-quality
observed color indices for the 2CD, the objects can not be classified using the
color-selection method.

\subsection{Gaia RP spectra\label{sec:rpspectra}}

The Gaia DR3 LPV (\citealt{lebzelter2022}) data provide the low-resolution RP
spectra at visual wavelength bands to identify C-rich stars. Using the Gaia RP
spectral data for the cross-identified 57,482 OGLE4 Mira variables (see
Table~\ref{tab:tab1}), they can be classified into 56,596 O-rich and 886 C-rich
stars by using the selection criteria (7 $<$ median\_delta\_wl\_rp $<$ 11 and
G$_{BP}$ $<$ 19 mag) presented in Section 2.4 in \citet{lebzelter2022}. The lower
panel of Figure~\ref{f1} shows the WISE-2MASS 2CD for OAGB and CAGB stars
classified from the Gaia RP spectra for the OGLE4 Mira variables.

Figure~\ref{f2} shows number density distributions of the Gaia magnitudes at the
BP band for OAGB and CAGB stars classified from the Gaia RP spectra for the OGLE4
Mira variables. The number density in the histogram denotes the raw count divided
by the total number of counts and the bin width, so that the area under the
histogram integrates to one. We find that most CAGB stars classified from the
Gaia RP spectra are bright nearby AGB stars.

We find that the classification using the Gaia RP spectra is right for a portion
of known AGB stars with IR spectroscopic data. Compared with the color-selection
method (see Section \ref{sec:w2cd}), we find that the accuracy ratio of the RP
spectral method is lower (see Section \ref{sec:agb-iras}). This could be due to
severe interstellar extinction processes at the visual bands or the effect of
dust envelopes around AGB stars. The RP spectral data would be more useful for
studying nearby AGB stars with thin dust envelopes.

In this work, we do not use the Gaia RP spectral data for classifying the OGLE4
Mira variables because we believe that the color-selection method (see Section
\ref{sec:w2cd}) is more useful for the large sample of AGB stars in our Galaxy.
However, the catalog data for all of the sample AGB stars (see
Section~\ref{sec:newagb}) will include the information from the available Gaia RP
spectral data.

\subsection{Period-magnitude and period-color relations\label{sec:pmr}}

Figure~\ref{f3} shows period-magnitude relations at W2[4.6] and W3[12] bands for
the OGLE4 Mira variables from \citet{iwanek2022} with known distances from Gaia
DR3. For each object, the absolute magnitudes at the two WISE bands are
calculated from the corrected distances from \citet{bailer-jones2021} for the
cross-matched Gaia DR3 LPV object (\citealt{lebzelter2022}). Generally, we find
that Mira variables with longer periods show brighter IR fluxes. The large
deviations from the linear relations would be mainly due to uncertainty in the
Gaia distances. As expected for Mira variables (e.g., \citealt{swu13};
\citealt{suh2020}), we find that the OGLE4 Mira variables occupy a single
sequence on the period-magnitude diagrams.

Figure~\ref{f4} shows period-color relations for 59,928 OAGB and 2674 CAGB stars
classified from the 65,981 OGLE4 Mira variables (see Section~\ref{sec:w2cd}).
Generally, we find that Mira variables with longer periods show redder IR colors.
Mira variables with longer periods are expected to have thicker dust envelopes as
they are generally more evolved AGB stars.

\subsection{IRAS PSC sources\label{sec:iras}}

Because IRAS has very low angular resolution, the AllWISE or 2MASS counterparts
obtained from the positions of cross-matched AKARI or MSX sources (with a higher
angular resolution; see Section~\ref{sec:photdata}) would be much more reliable.

For all of the 245,889 IRAS PSC sources, we have found the AKARI, MSX, AllWISE,
2MASS, OGLE4, Gaia DR3, and AAVSO counterparts by using following method. We
cross-identify the AKARI PSC, MSX, and AKARI BSC counterpart by finding the
nearest source within 60$\arcsec$ using the position given in the IRAS PSC
(version 2.1). Then, we cross-identify the AllWISE, 2MASS, and OGLE4 counterparts
using the best position of the available AKARI PSC, MSX, or AKARI BSC
counterpart. Only when there is no AKARI or MSX counterpart, we use the position
of the IRAS PSC.

The similar method was used in \citet{suh2018} and \citet{suh2021}. But in this
work, we use the MSX data in addition to the AKARI data to improve the accuracy
of the cross-match (see Section ~\ref{sec:crossmatch}).

From the 65,981 OGLE4 Mira variables, we find 18,413 IRAS counterparts (excluding
duplicate sources) within 60$\arcsec$. From the 18,426 sources, we find that
16,379 sources are positive (or reliable) counterparts, for which the AllWISE
counterpart of the IRAS source is the same as the AllWISE counterpart of the
OGLE4M object (see Table~\ref{tab:tab2}).

\section{A new sample of AGB stars\label{sec:newagb}}

\citet{suh2021} presented a catalog AGB stars in our Galaxy in two parts: one is
based on the IRAS PSC and the other one is based on the AllWISE source catalog.
In the IRAS based catalog, there are 5908 OAGB-IRAS and 3596 CAGB-IRAS objects.
And in the AllWISE based catalog, there are 5301 OAGB-WISE and 3576 CAGB-WISE
objects. Because IRAS has lower angular resolution and weaker sensitivity (see
Section~\ref{sec:photdata}) than WISE, the AGB-IRAS objects are generally
brighter than the AGB-IRAS objects.

In this work, we revise and update the catalog using the new sample of OGLE4 Mira
variables presented by \citet{iwanek2022}. We use the color-selection method (see
Section \ref{sec:w2cd}) to classify the OGLE4 Mira variables into OAGB or CAGB
stars, except for the objects for which more reliable IR spectroscopic data are
available.

We present a new catalog of 19,196 OAGB-IRAS, 4118 CAGB-IRAS, 45,413 OAGB-WISE,
and 5366 CAGB-WISE objects (see Tables~\ref{tab:tab3} and \ref{tab:tab4}). For
the new list of AGB stars, we cross-identify the IRAS, AKARI, MSX, AllWISE,
2MASS, OGLE4, Gaia DR3, and AAVSO counterparts and present the new catalog data,
which contain all of the available information.

Among the 65,981 OGLE4 Mira variables from \citet{iwanek2022}, 62,989 objects
(15,248 OAGB-IRAS, 797 CAGB-IRAS, 44,977 OAGB-WISE, and 1967 CAGB-WISE) are
included in the new catalog of OAGB and CAGB stars and 2000 objects are
classified into NI-O4 or NW-O4 subgroups (see Tables~\ref{tab:tab3} and
\ref{tab:tab4}). And 992 objects are not included in the catalog because 984
objects do not have the AllWISE counterparts and can not be cross-identified as
any known AGB stars and another 8 objects can be cross-identified as known S type
or silicate-carbon stars.

\begin{table*}
\centering
\caption{IRAS PSC sources\label{tab:tab2}}
\begin{tabular}{llllllllll}
\toprule
IRAS PSC$^1$ & AKARI PSC$^2$ & MSX$^2$ & AKARI BSC$^2$ & AllWISE$^3$ & 2MASS$^3$ & OGLE4M$^{3,4}$ & Gaia$^{3,5}$\\
\midrule
245,889  & 159,031 & 82,001 & 61,705 & 245,756 & 245,049 &  16,379 & 98,522 \\
\bottomrule
\end{tabular}
\tabnote{
$^1$version 2.1.
$^2$The number of counterparts within 60$\arcsec$ using the IRAS position (duplicate sources are exlcuded).
$^3$The number of counterparts within 60$\arcsec$ using the best position from IRAS, AKARI, and MSX (duplicate sources are exlcuded).
$^4$The number of positive counterparts within 60$\arcsec$ (see Section~\ref{sec:iras}).
$^5$Gaia DR3 LPV object (\citealt{lebzelter2022}).}
\end{table*}

\begin{table*}
\centering
\caption{Sample of AGB stars based on the IRAS PSC (AGB-IRAS) \label{tab:tab3}}
\begin{tabular}{llllllll}
\toprule
Class     &Subgroup & Reference & Number & Selected & Duplicate & AE$^1$ & Remaining \\
\midrule
OAGB-IRAS & OAGB-IRAS(2021) &\citet{suh2021} & 5908  & 5908 & 0  & -1$^2$ & 5907 \\
OAGB-IRAS & OI-O4S &This work & 16,379$^3$ & 2273$^4$  & 646$^5$ & 0 & 1627 \\
OAGB-IRAS & OI-O4C &This work & 16,379$^3$ & 12,920$^6$  & 1258$^{5}$  & 0 & 11,662 \\
OAGB-IRAS & OAGB-IRAS(2022) &Total & - & - & - & - & 19,196   \\
\midrule
CAGB-IRAS & CAGB-IRAS(2021) &\citet{suh2021} &  3596  & 3596 & 0  & -1$^{7}$ & 3595 \\
CAGB-IRAS & CI-O4S & This work & 16,379$^3$ & 295$^8$ & 243$^9$  & 0  & 52 \\
CAGB-IRAS & CI-O4C & This work & 16,379$^3$ & 493$^6$ & 22$^{10}$ & 0 & 471 \\
CAGB-IRAS & CAGB-IRAS(2022) &Total & -  & - & -  & - & 4118 \\
\midrule
No class & NI-O4 & This work  & 16,379$^3$ & 398$^6$  & 37$^5$ & - & 361 \\
\bottomrule
\end{tabular}
\tabnote{
$^1$added or excluded.
$^2$IRAS 17056-3959 is a carbon star (now in CI-O4S).
$^3$the number (16379) of positive IRAS counterparts from all of the OGLE4 Mira variables (see Section~\ref{sec:iras}).
$^4$objects with O-rich spectra from IRAS LRS and/or SIMBAD.
$^5$in OAGB-IRAS(2021) or CAGB-IRAS(2021) or in the list of S stars or silicate-carbon stars in AGB-IRAS (\citealt{suh2021}).
$^6$color-selected OAGB, CAGB, or No class objects (see Section~\ref{sec:w2cd}).
$^7$IRAS 17514-3354 is an OAGB star (now in OI-O4C).
$^8$objects with C-rich spectra from SIMBAD.
$^9$in CI-2021.
$^{10}$in CI-2021 or OI-2021.}
\end{table*}

\begin{table*}
\centering
\caption{Sample of AGB stars based on the AllWISE source catalog (AGB-WISE) \label{tab:tab4}}
\begin{tabular}{llllllll}
\toprule
Class     &Subgroup & Reference & Number & Selected & Duplicate & Remaining \\
\midrule
OAGB-WISE & OAGB-WISE(2021) &\citet{suh2021} & 5301  & 5301 & 4$^1$+49$^2$   & 5248 \\
OAGB-WISE & OW-O4C &This work & 64,962$^3$   & 59,928$^4$  & 14,936$^5$+4827$^6$  & 40,165 \\
OAGB-WISE & OAGB-WISE(2022) &Total & - & - & -   & 45,413 \\
\midrule
CAGB-WISE & CAGB-WISE(2021) &\citet{suh2021} &  3576  & 3576 & 2$^7$+2$^8$   & 3572 \\
CAGB-WISE & CW-O4C &This work & 64,962$^3$ & 2674$^{4}$ & 728$^9$+152$^6$   & 1794 \\
CAGB-WISE & CAGB-WISE(2022) &Total & -  & - & -  & 5366 \\
\midrule
No class & NW-O4 &This work  & 64,962$^3$ & 2360$^{4}$ & 715$^{10}$+6$^6$  & 1639 \\
\bottomrule
\end{tabular}
\tabnote{
$^1$in OAGB-IRAS(2021).
$^2$in OI-O4S or OI-O4C.
$^3$the number of OGLE4 Mira variables with AllWISE counterparts (see Table~\ref{tab:tab1}).
$^4$color-selected OAGB, CAGB, or No class objects (see Section~\ref{sec:w2cd}).
$^5$in OI-O4S, OI-O4C, or CI-O4S.
$^6$in OAGB-WISE(2021) or CAGB-WISE(2021).
$^7$in CAGB-IRAS(2021) or OAGB-IRAS(2021).
$^8$in CI-O4S or CI-O4C.
$^9$in CI-O4S, CI-O4C, or OI-O4S.
$^{10}$in NI-O4.}
\end{table*}

\subsection{AGB stars based on the IRAS PSC\label{sec:agb-iras}}

\citet{suh2021} presented a catalog AGB stars in our Galaxy that is based on the
IRAS PSC: 5908 OAGB-IRAS and 3596 CAGB-IRAS objects. We revise and update the
catalog using the new sample of OGLE4 Mira variables presented by
\citet{iwanek2022}.

In identifying important dust features of AGB stars, IRAS Low Resolution
Spectrograph (LRS; $\lambda$ = 8$-$22 $\mu$m) data are useful
(\citealt{kwok1997}). For OAGB stars, the IRAS LRS class E objects show the 10
$\mu$m silicate feature in emission and class A objects show the 10 $\mu$m
silicate feature in absorption. For CAGB stars, the LRS class C objects show the
11.3 $\mu$m SiC feature in emission.

For CAGB stars in the IRAS LRS class C, the accuracy ratio of the color-selection
method (see Section \ref{sec:w2cd}) is 82 \%, while the accuracy ratio of the
Gaia RP spectral method (see Section \ref{sec:rpspectra}) is 55 \%. For OAGB
stars in the IRAS LRS class E or A, the inaccuracy ratios of the two methods are
about the same.

Table \ref{tab:tab3} lists a new catalog of AGB-IRAS objects. We correct some
misidentified objects from \citet{suh2021}. We add the OGLE4 Mira variables
(presented by \citealt{iwanek2022}) with the positive IRAS counterparts (see
Section~\ref{sec:iras}), which are classified into OAGB and CAGB stars using the
color-selection method using the WISE-2MASS 2CD (see Section \ref{sec:w2cd}).

In classifying into OAGB or CAGB, the more reliable method using the spectral
information (the spectral type from the IRAS LRS or SIMBAD) gets priority over
the color-selection method using IR 2CDs.

Finally, we present a new list of 23,314 AGB-IRAS objects, which are classified
into 19,196 OAGB-IRAS and 4118 CAGB-IRAS stars.

\subsection{AGB stars based on the AllWISE source catalog\label{sec:agb-wise}}

\citet{suh2021} presented a catalog AGB stars in our Galaxy that is based on the
AllWISE source catalog: 5301 OAGB-WISE and 3576 CAGB-WISE objects. We revise and
update the catalog using the new sample of OGLE4 Mira variables presented by
\citet{iwanek2022}.

Table \ref{tab:tab4} lists a new catalog of AGB-WISE objects. They are identified
based on the AllWISE source catalog and do not have positive IRAS counterparts.
We make corrections for some misidentified AGB-WISE objects from \citet{suh2021}.
We add the OGLE4 Mira variables (presented by \citealt{iwanek2022}) with the
AllWISE counterparts.

Now, we present a new list of 50,779 AGB-WISE objects, which are classified into
45,413 OAGB-WISE and 5366 CAGB-WISE stars.

\begin{figure*}[!t]
\centering
\includegraphics[width=120mm]{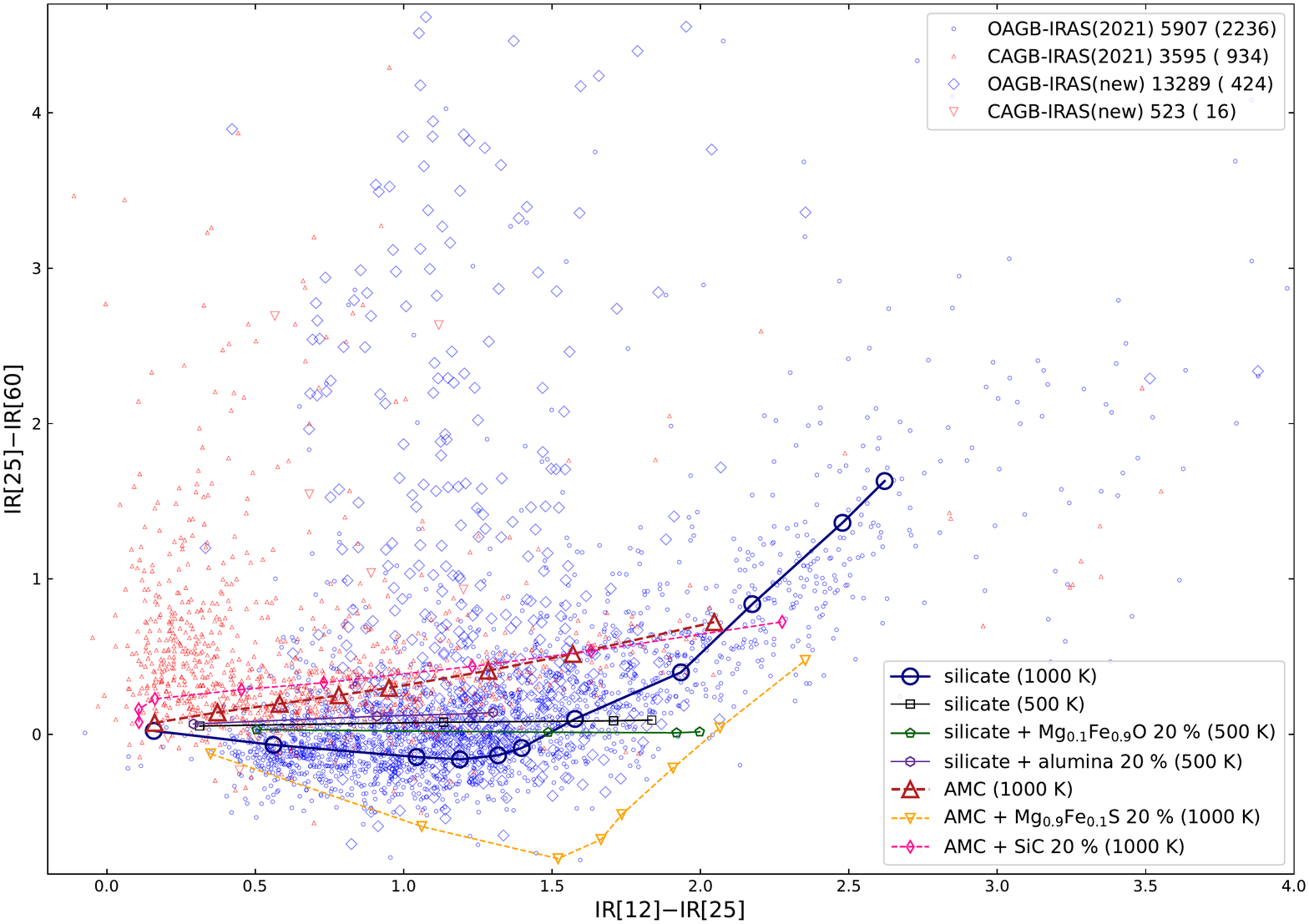} \\
\includegraphics[width=120mm]{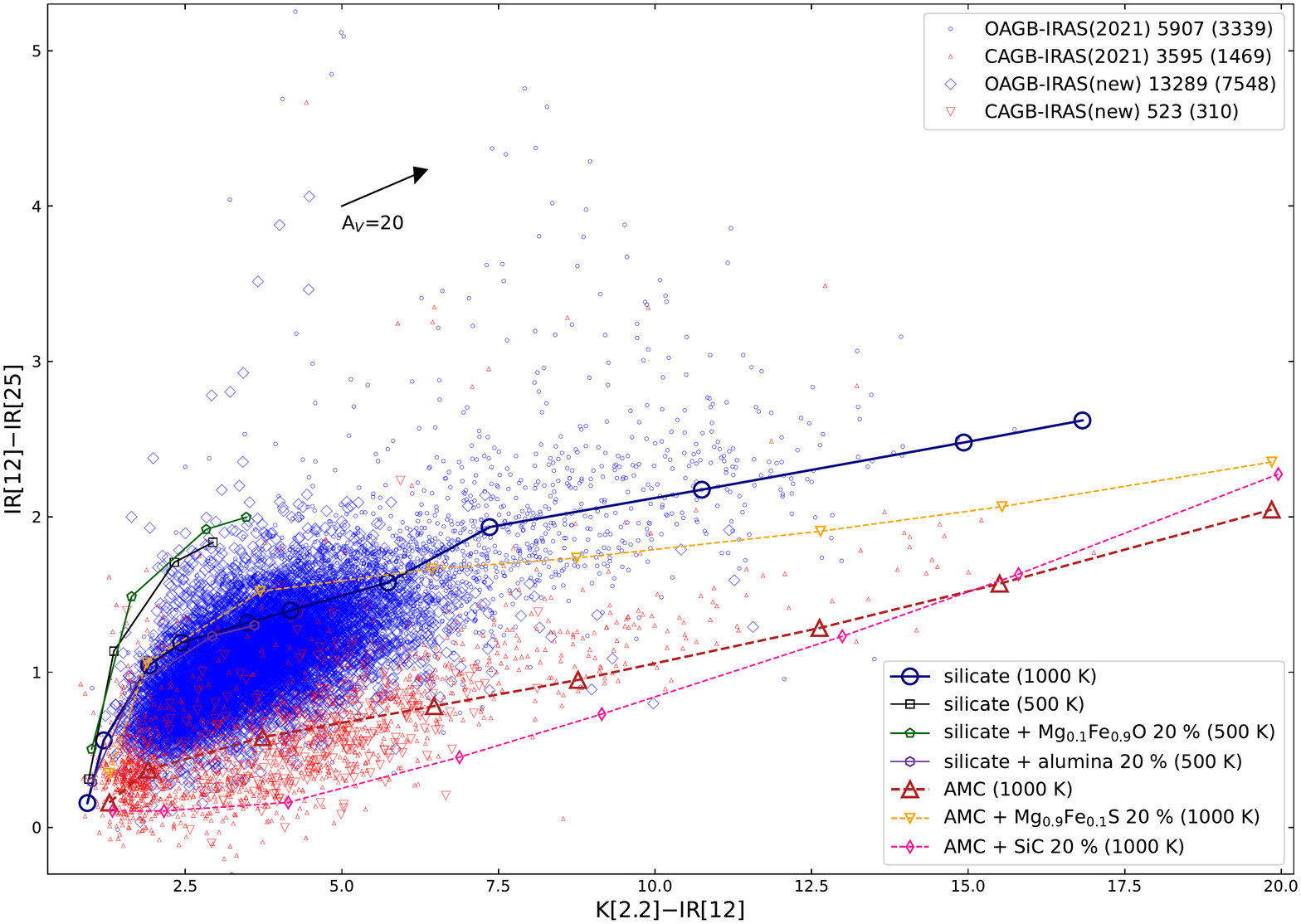}
\caption{IRAS-2MASS 2CDs for AGB-IRAS stars in the previous catalog of \citet{suh2021} and new AGB-IRAS stars identified from the OGLE4 Mira variables
(see Tables~\ref{tab:tab3}) compared with theoretical models (see Section~\ref{sec:models}).
The number of objects and the number of plotted objects (in parenthesis) with good-quality observed data are shown for each class.
\label{f5}}
\end{figure*}

\begin{figure*}[!t]
\centering
\includegraphics[width=120mm]{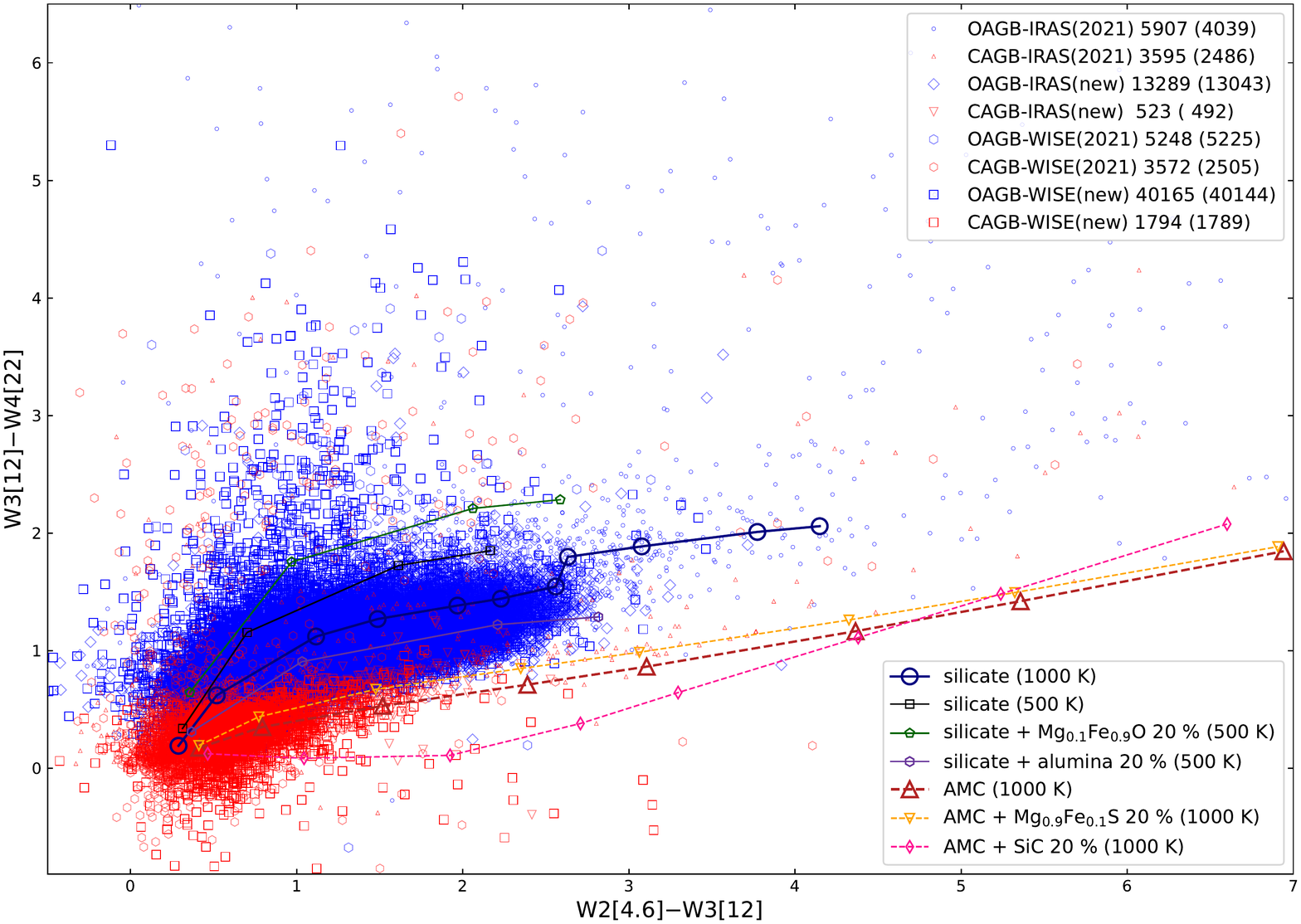} \\
\includegraphics[width=120mm]{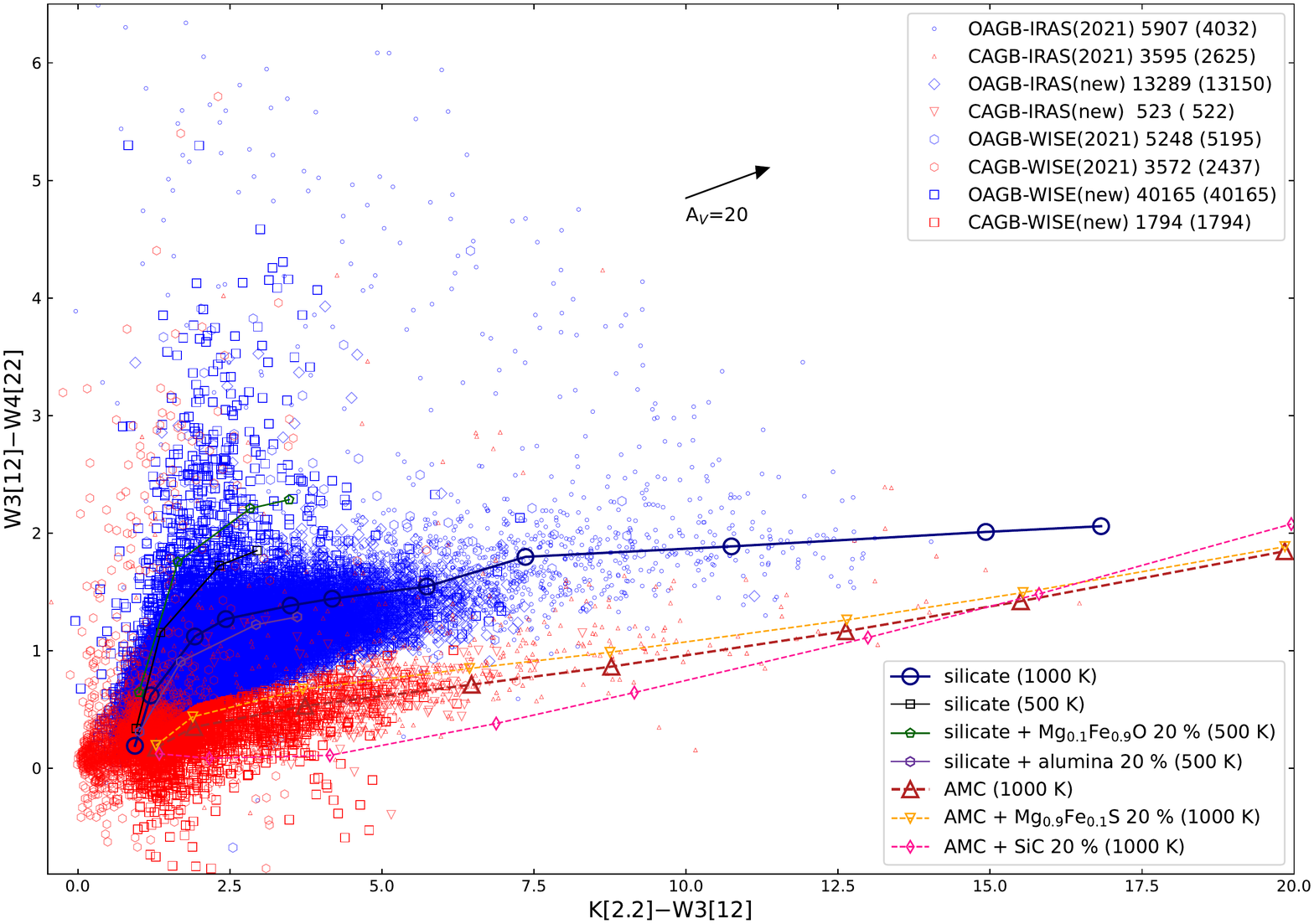}
\caption{WISE-2MASS 2CDs for all AGB stars in the previous catalog of \citet{suh2021} and new AGB stars identified from the OGLE4 Mira variables
(see Tables~\ref{tab:tab3} and \ref{tab:tab4}) compared with theoretical models (see Section~\ref{sec:models}).
The number of objects and the number of plotted objects (in parenthesis) with good-quality observed data are shown for each class.
\label{f6}}
\end{figure*}

\begin{figure*}[!t]
\centering
\includegraphics[width=140mm]{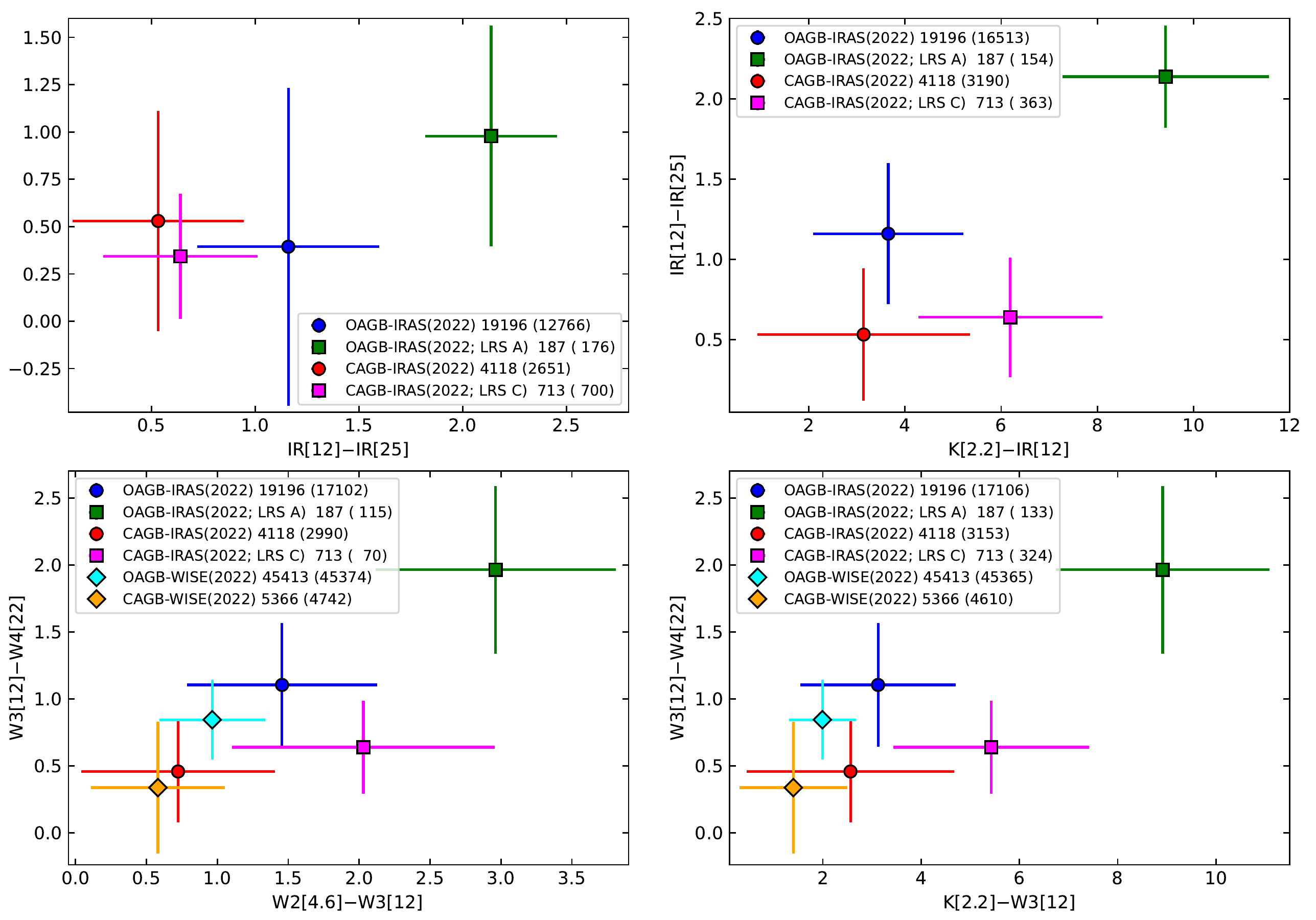}
\caption{Averaged observed IR colors for different classes of AGB stars
(see Tables~\ref{tab:tab3} and \ref{tab:tab4}; Figures~\ref{f4} and~\ref{f5}).
The number of objects and the number of considered objects (in parenthesis) with good-quality observed data for the horizontal axis are shown for each class.
\label{f7}}
\end{figure*}

\begin{figure}[!t]
\centering
\includegraphics[width=86mm]{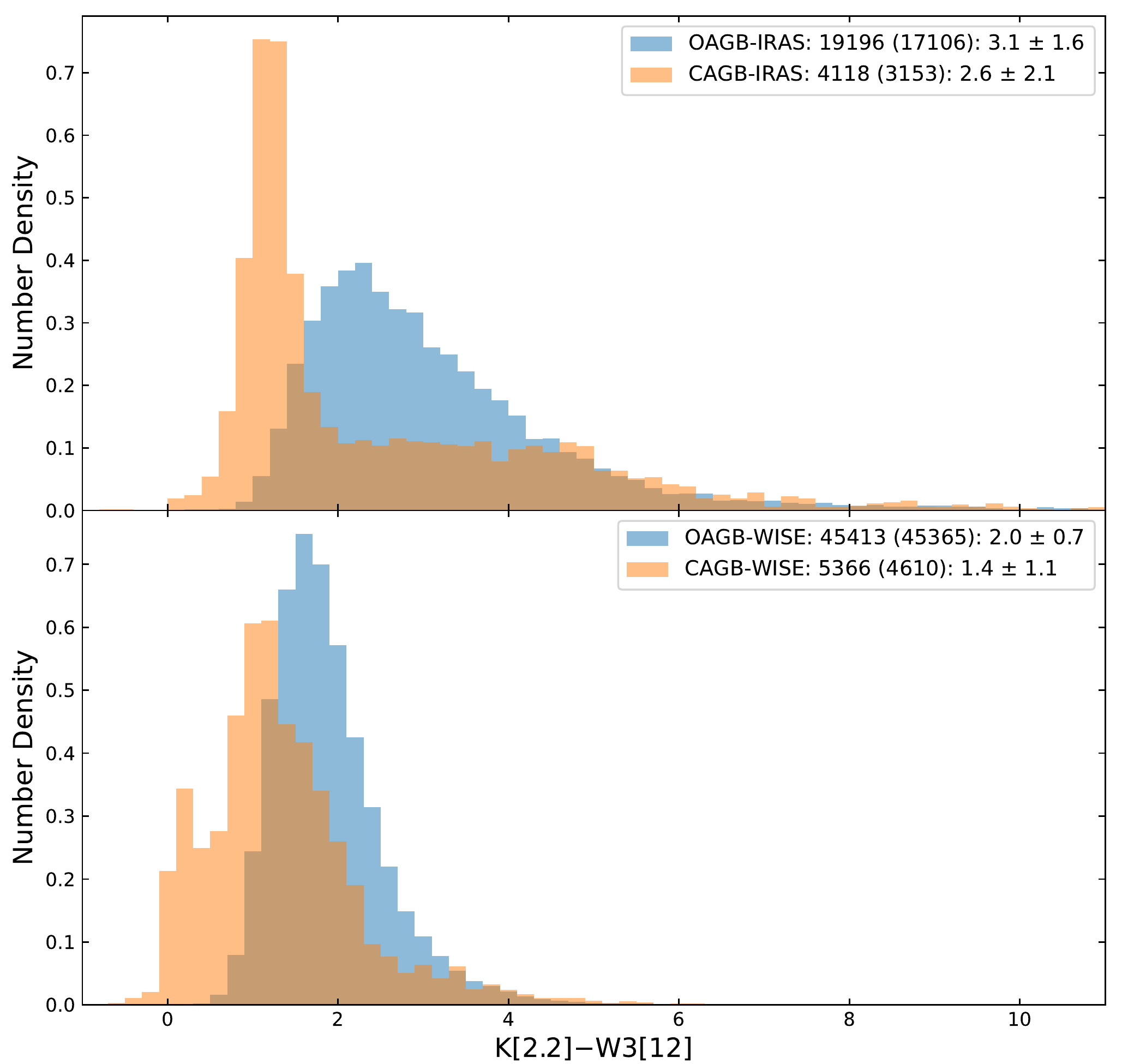}
\caption{Number density distributions of observed K[2.2]$-$W3[12] indices for AGB-IRAS and AGB-WISE stars.
The number of objects and the number of considered objects (in parenthesis) with good-quality observed data are shown for each class.
\label{f8}}
\end{figure}

\begin{figure}[!t]
\centering
\includegraphics[width=86mm]{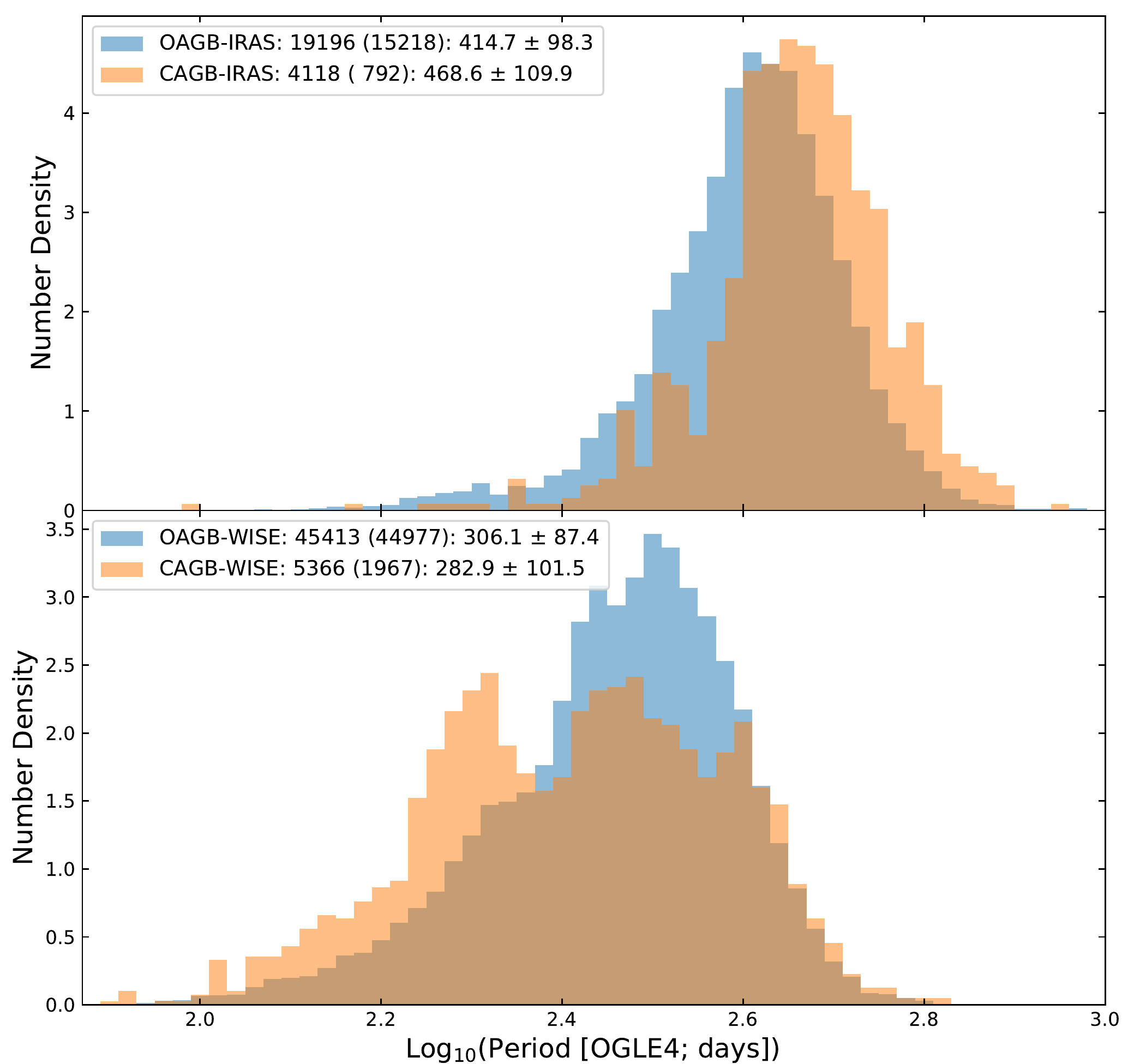}
\caption{Number density distributions of the observed OGLE4 periods for AGB-IRAS and AGB-WISE stars.
\label{f9}}
\end{figure}

\section{IR 2CDs and Discussion\label{sec:theory}}

Figures \ref{f5} and \ref{f6} show various 2CDs for AGB stars in the new revised
and extended catalog (see Section~\ref{sec:newagb}; Tables~\ref{tab:tab3} and
\ref{tab:tab4}). On all of the 2CDs in Figures \ref{f1}, \ref{f5}, and \ref{f6},
theoretical models for AGB stars are also plotted to be compared with the
observations.

\subsection{Theoretical Dust Shell Models\label{sec:models}}

In this paper, we use the same theoretical dust shell models as the ones used in
\citet{suh2020} and \citet{suh2021} (see Section 3 and Table 7 in \citet{suh2021}
for the summary). We use the radiative transfer code DUSTY
(\citealt{ivezic1997}), which assumes a spherically symmetric dust shell around a
central star. We assume that the central stars is a black body (T = 3000-2000 K).

For all models, we use a simple power law ($\rho \propto r^{-2}$) for the dust
density distribution and assume that the dust formation temperature ($T_c$) is
1000-500 K. The outer radius of the dust shell is taken to be $10^4$ times the
inner radius, the fiducial wavelength of the dust shell optical depth
($\tau_{10}$) is 10 $\mu$m, and the uniform radius of spherical dust grains is
0.1 $\mu$m.

For OAGB stars, we use dust optical constants of warm and cold silicates
(\citealt{suh1999}), amorphous alumina (\citealt{suh2016}), and
Fe$_{0.9}$Mg$_{0.1}$O (\citealt{henning1995}). For CAGB stars, we use dust
optical constants of AMC (\citealt{suh2000}), SiC (\citealt{pegouri1988}), and
Mg$_{0.9}$Fe$_{0.1}$S (\citealt{begemann1994}).

On the 2CDs in Figures \ref{f1}, \ref{f4}, and \ref{f5}, theoretical models for
AGB stars are plotted. For OAGB models (silicate $T_c$ = 1000 K), the model
results for $\tau_{10}$ = 0.001, 0.01, 0.05, 0.1, 0.5, 1, 3, 7, 15, 30, and 40
are plotted from left to right on the 2CDs. For CAGB models (AMC $T_c$ = 1000 K),
the model results for $\tau_{10}$ = 0.001, 0.01, 0.1, 0.5, 1, 2, 3, and 5 are
plotted from left to right on the 2CDs.

For physical implications of the theoretical dust shell models, refer to
\citet{suh2020} and \citet{suh2021}. \citet{suh2020} investigated IR properties
of AGB Stars in Our Galaxy and the Magellanic Clouds and discussed the physical
implications (e.g., mass-loss rates, masses, ages, and metallicities) of the
theoretical model parameters for different classes of AGB stars. Those model
parameters were obtained from the comparison of the theoretical dust shell models
with the observations on various IR 2CDs.

\subsection{Comparison between Theory and Observations\label{sec:comparison}}

On various IR 2CDs in Figures~\ref{f5} and~\ref{f6}, we compare the observations
with the theoretical dust shell models for AGB stars. Generally, the theoretical
dust shell models for OAGB and CAGB stars can reproduce the observed points
fairly well.

The upper panel of Figure~\ref{f5} plots an IRAS 2CD using [25]$-$[60] versus
[12]$-$[25]. A group of CAGB stars in the upper-left region are visual carbon
stars that show excessive flux at 60 $\mu$m due to the remnant of the earlier
phase when they were OAGB stars (e.g. \citealt{ck1990}). Another group of CAGB
stars in the lower region, which extends to the right side, are IR carbon stars.
The IR carbon stars with redder [12]$-$[25] colors are believed to be more
evolved or massive stars with thicker dust envelope (see Section 1 in
\citealt{suh2020} for origin and evolution of carbon stars).

The lower panel of Figure~\ref{f5} plots an 2MASS-IRAS 2CD using IR[12]$-$IR[60]
versus K[2.2]$-$IR[12]. On this 2CD, the distinguishment between OAGB and CAGB
stars is clearer. Visual carbon stars are located in the lower-left region on
this 2CD, because they show bluer K[2.2]$-$IR[12] and IR[12]$-$IR[60] colors than
IR carbon stars.

Figure~\ref{f6} shows 2CDs using WISE and 2MASS colors. The upper panel of
Figure~\ref{f6} shows a WISE 2CDs using W3[12]$-$W4[22] versus W2[4.6]$-$W3[12].
The lower panel of Figure~\ref{f6} shows a WISE-2MASS 2CD using W3[12]$-$W4[22]
versus K[2.2]$-$W3[12]. On these 2CDs, the theoretical dust shell models for CAGB
stars with various C-rich dust grains can reproduce a wider range of observed
W3[12]$-$W4[22] colors.

In the lower regions of the 2CDs in Figure~\ref{f6}, we find that there are many
newly identified CAGB-IRAS or CAGB-WISE objects in this work, which are
color-selected from the OGLE4 Mira variables (see Section \ref{sec:w2cd}). We may
need to confirm the C-rich nature of these objects by using a more reliable IR
spectroscopic method. The objects in the upper-left regions on the 2CDs in
Figure~\ref{f6} are likely to be early stage OAGB stars or visual carbon stars
with thin and detached dust shells ($T_c$ = 200-300 K; see \citealt{suh2020}).

\subsection{Statistical properties of AGB stars\label{sec:statisics}}

Figure~\ref{f7} shows the error bar plots of the observed colors on the various
2CDs used for all classes of AGB stars presented in Figures~\ref{f5}
and~\ref{f6}. Because IRAS LRS class A objects (mostly OH/IR stars) and class C
objects (IR carbon stars) have thicker dust envelopes, they are believed to be
more evolved or massive AGB stars. We find that these more evolved OAGB or CAGB
objects with thicker dust envelopes are show redder colors for all of the IR
color indices except for the [25]$-$[60] color. This is because visual carbon
stars with thin dust envelopes show excessive flux at 60 $\mu$m. Generally,
AGB-WISE objects show bluer colors compared with AGB-IRAS objects. This would be
because most AGB-WISE objects are less evolved and have thinner dust shells than
AGB-IRAS objects.

Figure~\ref{f8} shows number density distributions of K[2.2]$-$W3[12] indices for
AGB-IRAS and AGB-WISE stars. Compared with AGB-WISE objects, more AGB-IRAS
objects show redder colors because they have thicker dust envelopes. Compared
with CAGB-IRAS objects, OAGB-IRAS objects show redder colors.

Figure~\ref{f9} shows number density distributions of the OGLE4 periods for
AGB-IRAS and AGB-WISE stars. We find that more AGB-IRAS objects show longer
averaged pulsation periods than AGB-WISE objects, which are generally less
evolved or less massive. We also find that CAGB-IRAS stars show generally longer
periods than OAGB-IRAS stars. This would be due to a selection effect of OGLE4
observations. Because the OGLE4 periods are obtained from the $I$ filter
photometry, most OAGB-IRAS stars with thick dust envelopes and longer pulsation
periods (OH/IR stars) are not observed by the OGLE4 observations.

\section{Summary\label{sec:summary}}

We have investigated infrared properties of OGLE4 Mira variables in our Galaxy.
For each object, we have cross-identified the AllWISE, 2MASS, and IRAS
counterparts. We have presented various infrared two-color diagrams (2CDs) and
period-magnitude and period-color relations for the Mira variables. Generally,
the Mira variables with longer periods are brighter in the IR fluxes and redder
in IR colors.

We have revised and updated the catalog of AGB stars in our Galaxy from
\citet{suh2021} using the new sample of OGLE4 Mira variables. Now, we present a
new catalog of 74,093 (64,609 OAGB and 9484 CAGB) stars in our Galaxy. A group of
23,314 (19,196 OAGB-IRAS and 4118 CAGB-IRAS) stars are identified based on the
IRAS PSC and another group of 50,779 (45,413 OAGB-WISE and 5366 CAGB-WISE) stars
are identified based on the AllWISE source catalog.

For all of the AGB stars in the new catalog, we have cross-identified the IRAS,
AKARI, MSX, AllWISE, 2MASS, OGLE4, Gaia DR3, and AAVSO counterparts and presented
a new catalog data, which contains all of the available information.

We have presented various IR 2CDs for all AGB stars in the catalog. Comparing the
observations with the theory on the 2CDs, we have found that basic theoretical
dust shell models can account for the IR observations fairly well for most of the
AGB stars.

We have found that the AGB-IRAS objects show longer averaged pulsation periods
and redder averaged IR colors than AGB-WISE objects, which are generally less
evolved or less massive.

The new revised and updated data for the catalog of AGB stars in our Galaxy
(version 2022) will be accessible through the author's webpage
(\url{https://web.chungbuk.ac.kr/~kwsuh/agb.htm}).

\acknowledgments

This research was supported by Basic Science Research Program through the National Research
 Foundation of Korea (NRF) funded by the Ministry of Education (2022R1I1A3055131).
 This research has made use of the VizieR catalogue access tool and the SIMBAD database, operated at CDS, Strasbourg, France.
 This research has made use of the NASA/IPAC Infrared Science Archive, which is operated
 by the Jet Propulsion Laboratory, California Institute of Technology, under contract with
 the National Aeronautics and Space Administration.

%-------------------------------------------------------------------
\end{document}